\documentclass{elsart}
\usepackage{graphicx,amssymb}
\journal{Physica A}
\begin{document}
\begin{frontmatter}

\title{Mesoscopic constitutive relations for dilute polymer solutions}
\author{C. M\'{a}laga\thanksref{fn1}},
\author{F. Mandujano},
\author{I. Santamar\'{i}a-Holek}

\thanks[fn1]{E-mail: camalaga@yahoo.co.uk}

\address{Facultad de Ciencias, Universidad Nacional Aut\'{o}noma
de M\'{e}xico.\\ Circuito exterior de Ciudad Universitaria. 04510,
D. F., M\'{e}xico}

\begin{abstract}
A novel approach to the dynamics of dilute solutions of polymer
molecules under flow conditions is proposed by applying the rules of
mesoscopic nonequilibrium thermodynamics (MNET). The probability
density describing the state of the system is taken to be a function
of the position and velocity of the molecules, and on a local vector
parameter accounting for its deformation. This function obeys a
generalized Fokker-Planck equation, obtained by calculating the
entropy production of the system, and identifying the corresponding
probability currents in terms of generalized forces. In simple form,
this coarse-grained description allows one to derive hydrodynamic
equations where molecular deformation and diffusion effects are
coupled. A class of non-linear constitutive relations for the pressure
tensor are obtained. Particular models are considered and
compared with experiments.
\end{abstract}
\begin{keyword}
Fokker-Planck equations; FENE models; Polymer solutions; Non-Newtonian
Phenomena; Constitutive relations.
\end{keyword}
\end{frontmatter}

\section{Introduction}

The statistical study of chain models of polymer molecules
\cite{flory,bird71,warner} have led to a collection of constitutive
relations for the description of the elastic properties of polymer
solutions \cite{chilcott,rallison1}, known as FENE dumbbell models, in
which the restoring force depends on the end-to-end vector of the
chain. By avoiding phenomena related to the fine structure of the
polymer molecules, those models depict an intuitive picture of the
physics involved in the molecule deformation \cite{hinch}, and
reproduce non-Newtonian phenomena, such as birefringent pipes
\cite{harlen2}, negative wake \cite{harlen4} and cusp formation
\cite{malaga}.

Numerical simulations of molecular chain models have
shown that the FENE approximation is useful to describe systems under
flow conditions, for which the largest relaxation time of the polymer
dominates the dynamics of the molecule, but are inaccurate for
molecules under unsteady flow conditions in which molecules can
explore faster relaxation modes \cite{ghosh}. 
Although simulations predict a more realistic
restoring force, they are unable to
describe its evolution in terms of a simple mathematical formulation
for arbitrary flows, legitimizing the search for a simple
description which captures short time scale physics.

In this paper, a theoretical framework originally proposed in
Ref. \cite{rubiPNAS,rubi} has been adopted to cope with the main
features of dilute polymer suspensions under stationary flow.  A
coarse-grained description of the system dynamics is given in terms of
a Fokker-Planck equation for the probability density depending on the
parameters accounting for the motion of the polymer center of mass,
and a single vector parameter (called the distortion vector)
accounting for its elastic deformation. The Fokker-Planck equation can
be constructed after invoking probability conservation, calculating
the entropy production in the phase space of the system and following
the scheme of mesoscopic non-equilibrium thermodynamics
\cite{Perez94}. Moreover, it can be used to formulate a hydrodynamic
description in terms of the evolution equations for the moments of the
probability. Application to particular cases can be performed by
incorporating different models for the Onsager tensor coefficients
and the finite-size of the molecules \cite{checoloco2}.

The paper is organized as follows. In Sec. \textbf{2}, the derivation
of the generalized Fokker-Planck equation is presented. In
Sec. \textbf{3}, the evolution equations for the moments of the
probability are derived, and in Sec. \textbf{4}, are applied to
obtain constitutive relations for the pressure tensor in
particular cases. Sec. \textbf{5} is
devoted to summarize our main results.
            
\section{Fokker-Planck equations}

The system under study, a dilute polymer solution, will be regarded as
a polymer 'gas' suspended in a Newtonian solvent under stationary flow
conditions given by ${\bf v}_0({\bf r})$. The description of this 
system can be carried out in terms of a Fokker-Planck equation.
 
As a first approximation, the state of the polymer gas can be
characterized by the positions ${\bf r}$ and instantaneous velocities
${\bf u}$ of the center of mass of the polymer molecules
\cite{rubiPNAS,rubi,checoloco2,checoloco}, and by the vector ${\bf R}$
characterizing the gross distortion and orientation of the molecule
\cite{warner,hinch}. Here, it is assumed that the molecule segments
explore different configurations much faster than the variations in the
gross distortion due to the local velocity gradient, i.e. molecules
are distorted instantaneously with the fluid element in which they are
embedded. Notice that, unlike in the conventional
approaches \cite{warner}, here we introduce the velocity ${\bf u}$ as
an independent variable necessary to determine the nonequilibrium
state of the system at the conditions imposed by the velocity
gradient. In the dilute regime, the description of the state of the
system can be carried out in terms of the single molecule probability
density $f({\bf r},{\bf u},{\bf R},t)$, normalized to the number of
molecules and satisfying the continuity equation
\begin{equation}
\label{contf}
	\frac{\partial f}{\partial t} + \nabla \cdot ({\bf u}f) = -
	\frac{\partial}{\partial {\bf u}} \cdot \left[ f {\bf V}_u \right] -
	\frac{\partial}{\partial {\bf R}} \cdot \left[ f {\bf V}_R \right],
\end{equation}
where ${\bf V}_u$ and ${\bf V}_R$ are unknown streaming velocities
in ${\bf u}$ and ${\bf R}$ subspaces 
\cite{rubi,checoloco2,checoloco}. 

The Fokker-Planck equation describing the dynamics of the system
follows from the explicit expressions of ${\bf V}_u$ and ${\bf V}_R$.
These can be obtained using the MNET formalism which introduces the Gibbs entropy
postulate \cite{mazur,rubiPNAS,rubi}
\begin{equation}
\label{gibbspost}
	s(t)= -k_B \int C \ln{\frac{f}{f_{leq}}} d{\bf u}d{\bf
	R} + s_{leq},
\end{equation}
where $k_B$ is the Boltzmann constant, $s$ and $s_{leq}$ are the 
entropy per mass unit and the entropy at local equilibrium of a given
reference state, respectively. $C \equiv
mf/ \rho$ represents the mass fraction of polymer molecules with a 
given ${\bf u}$ and ${\bf R}$, $m$ the molecule mass and $\rho$ the 
mass density field, given by
\begin{equation}
\label{ro}
	\rho({\bf r},t) = \int mf d{\bf u}d{\bf R}.
\end{equation}
The probability density $f_{leq}$ characterizing the
system at the reference state is given by
\begin{equation}
\label{fleq}
	f_{leq}({\bf r},{\bf u},{\bf R})=e^{\frac{m}{k_BT}\left[\mu_B -
	\frac{1}{2}\left({\bf u}-{\bf v}_0 \right)^2 - {\bf F} \cdot {\bf R} 	
\right]},
\end{equation}
where $T$ is the temperature of the heat bath, $\mu_B$ is the local 
equilibrium chemical potential and $\frac{1}{2}({\bf u}-{\bf v}_0)^2$ 
the kinetic energy of diffusion per unit mass. The last term in the 
argument of the exponential function, $ {\bf F} \cdot {\bf R}$, is the work 
exerted by the bath on the molecule in order to attain a distortion 
${\bf R}$. Here ${\bf F}$ is the restitutive force per unit mass of the 
molecule thought as a springy body. 

The Gibbs entropy postulate (\ref{gibbspost}) is consistent with the 
Gibbs equation \cite{groot,rubi}
\begin{equation}
\label{gibbseqn}
	T \delta s = \delta e - p \delta \rho^{-1} -
	\int \mu \delta C d{\bf u}d{\bf R},
\end{equation}
where $e$ is the internal energy per unit mass, $p$ is the pressure
and $\mu$ the nonequilibrium chemical potential. The last term of Eqn. 
(\ref{gibbseqn}) is reminiscent of the one corresponding to a mixture 
in which the components are specified by two continuum indexes 
(${\bf u}$ and ${\bf R}$). Eqns. (\ref{contf})-(\ref{gibbseqn}) can be 
used to obtain the explicit expressions for the streaming velocities 
${\bf V}_u$ and ${\bf V}_R$, by calculating the entropy production 
$\sigma$ of the system in the 
$\left({\bf r},{\bf u},{\bf R}\right)$-space, and taking
into account the evolution equation for $C$ \cite{checoloco}. The mass 
continuity equation is given by
\begin{equation}
\label{cont}
	\frac{d \rho}{d t} = - \rho \nabla \cdot {\bf v},
\end{equation}
where ${\bf v}({\bf r},t)$ is the average velocity field of the
molecules defined by 
\begin{equation}
\label{v}
	{\bf v}({\bf r},t)= \int {\bf u} C d{\bf u}d{\bf R},
\end{equation}
and the convective derivative given by $\frac{d}{dt} =
\frac{\partial}{\partial t} + {\bf v} \cdot \nabla$.
After taking the time derivative of the Gibbs equation (\ref{gibbseqn})
and integrating by parts, assuming that probability fluxes vanish at 
the boundaries, one obtains the entropy balance equation
\begin{equation}
\label{balance}
	\rho \frac{ds}{dt} = -\nabla \cdot {\bf J}_s + \sigma.
\end{equation}
In writing this equation, we have identified the entropy flow
${\bf J}_s$
\begin{eqnarray}
\label{js}
        {\bf J}_s & \equiv & - k_B \int ({\bf u} -{\bf v}) f(\ln{f} -1)
	d{\bf u}d{\bf R} \\ \nonumber & - & \frac{1}{T} \int ({\bf u}
	-{\bf v})f{\bf F} \cdot {\bf R} d{\bf u}d{\bf R}
	- \frac{m}{2T} \int ({\bf u} -{\bf v})f ({\bf u} -{\bf v})^2
	d{\bf u}d{\bf R},
\end{eqnarray}
the entropy production 
\begin{eqnarray}
\label{sigma}
        \sigma & \equiv & - \frac{m}{2T} \int f {\bf V}_u \cdot
	\frac{\partial \mu}{\partial {\bf u}} d{\bf u}d{\bf R} -
	\frac{m}{2T} \int f {\bf V}_R \cdot \frac{\partial
	\mu}{\partial {\bf R}} d{\bf u}d{\bf R} \nonumber \\ & - &
	\frac{m}{2T} \int f ({\bf u} -{\bf v}) \cdot \nabla ({\bf u}
	-{\bf v}_0)^2 d{\bf u}d{\bf R},
\end{eqnarray}
and assumed that the elastic force ${\bf F}$ is independent of ${\bf 
r}$ and ${\bf u}$. The incorporation of ${\bf R}$ into the description, 
introduces elastic contributions to both the entropy flow and the 
entropy production in Eqns. (\ref{js}) and (\ref{sigma}). They represent a 
novel ingredient not considered in previous MNET analysis (see, for 
example, Ref. \cite{rubi,checoloco2,checoloco}). 

After taking variations of Eqn. (\ref{gibbspost}) and comparing with 
Eqn. (\ref{sigma}), one arrives at the explicit expression of the
nonequilibrium chemical potential
\begin{equation}
\label{mu}
        \mu \equiv \frac{k_B T}{m} \ln{f} + \frac{1}{2} ({\bf u} -{\bf
	v}_0)^2 + {\bf F} \cdot {\bf R}.
\end{equation}
Once obtained the entropy production $\sigma$ (Eqn. (\ref{sigma})) and 
identified the nonequilibrium chemical potential $\mu$ (Eqn. 
(\ref{mu})) conjugated to the probability density defined over the phase space of 
the system, one can assume linear relationships between fluxes and 
forces, of the form 
\cite{rubiPNAS,rubi,checoloco2,checoloco,mazur} 
\begin{eqnarray} 
	f{\bf V}_u &=& -f \mathbb{L}_{uu} \cdot \frac{\partial
	\mu}{\partial {\bf u}} -f \mathbb{L}_{uR} \cdot
	\frac{\partial\mu}{\partial {\bf R}} -f \mathbb{L}_{ur} \cdot
	({\bf u} -{\bf v}_0) \cdot \left(\nabla {\bf v}_0\right)^T, 
\label{ju} \\
	f{\bf V}_R &=& -f \mathbb{L}_{Ru} \cdot \frac{\partial
	\mu}{\partial {\bf u}} -f \mathbb{L}_{RR} \cdot
	\frac{\partial\mu}{\partial {\bf R}} -f \mathbb{L}_{Rr} \cdot
	({\bf u} -{\bf v}_0) \cdot  \left(\nabla {\bf v}_0\right)^T, 
\label{jr}
\end{eqnarray}
where $\left(\nabla {\bf a} \right)_{ij}=\frac{\partial a_j}{\partial 
x_i}$ and the Onsager coefficient matrices $\mathbb{L}_{ij}$ satisfy the
generalized Onsager relations \cite{rubi-dufty}. Notice that unlike the 
usual framework of nonequilibrium thermodynamics given in ${\bf 
r}$-space  \cite{groot}, in the MNET formalism, the linear relationships 
(\ref{jr}) are formulated in the $({\bf r},{\bf u},{\bf R})$-space 
\cite{mazur,rubiPNAS}. This assumption will lead, in general, to nonlinear 
constitutive relations in the physical space ${\bf r}$ after a contraction of 
the description \cite{checoloco2}. Finally,
substituting equations (\ref{ju}) and (\ref{jr}) into the continuity
equation (\ref{contf}), one obtains
\begin{eqnarray}
	\frac{\partial f}{\partial t} + \nabla \cdot ({\bf u}f) =
	\nonumber \\
	\frac{\partial}{\partial {\bf u}} \cdot \left[\mathbb{C}_1 \cdot ({\bf 
u} -{\bf
	v}_0)f + \frac{k_B T}{m} \mathbb{L}_{uu}
	\cdot \frac{\partial f}{\partial {\bf u}} + \mathbb{L}_{uR} \cdot
	\left( {\bf F} f + \frac{k_B T}{m}
	\frac{\partial f}{\partial {\bf R}} \right) \right] \nonumber \\
	+
	\frac{\partial}{\partial {\bf R}} \cdot \left[\mathbb{C}_2 \cdot ({\bf 
u} -{\bf
	v}_0) f + \frac{k_B T}{m} \mathbb{L}_{Ru}
	\cdot \frac{\partial f}{\partial {\bf u}} + \mathbb{L}_{RR} \cdot
	\left( {\bf F} f + \frac{k_B T}{m}
	\frac{\partial f}{\partial {\bf R}} \right) \right],
\label{fuckingplanck}
\end{eqnarray}
which constitutes the Fokker-Planck equation describing the evolution
of the system at mesoscopic level. We have also defined the tensors
\begin{equation}
	\mathbb{C}_1  \equiv  \mathbb{L}_{uu} + \mathbb{L}_{ur} \cdot \nabla 
{\bf v}_0 \hspace{1cm} \mbox{and}  \hspace{1cm}
	\mathbb{C}_2  \equiv  \mathbb{L}_{Ru} + \mathbb{L}_{Rr} \cdot \nabla 
{\bf v}_0.
\label{c1}
\end{equation}
From these relations it follows that, in general, the transport
coefficients may depend on the imposed velocity gradient, implying the
breaking of the fluctuation-dissipation theorem at mesoscopic level,
and incorporating non-Newtonian behavior 
\cite{checoloco2,checoloco}. 

\section{Dynamics of the molecular distortion under flow 
conditions.}

The equation describing the dynamics of the distortion of a single
molecule can be obtained averaging over ${\bf u}$ the Fokker-Planck
equation (\ref{fuckingplanck}). Assuming that the Onsager coefficients do
not depend on ${\bf u}$, one obtains the equation
\begin{equation}
	\frac{\partial g}{\partial t} + \nabla \cdot ({\bf v}_R g) = 
	\frac{\partial}{\partial {\bf R}} \cdot \left[\mathbb{C}_2 \cdot ({\bf 
v}_R -{\bf
	v}_0) g + \mathbb{L}_{RR} \cdot
	\left( {\bf F} g + \frac{k_B T}{m}
	\frac{\partial g}{\partial {\bf R}} \right) \right],
\label{gevola}
\end{equation}
which governs the evolution of $g({\bf r},{\bf R},t) = \int f d{\bf
u}$, that represents the probability of finding a molecule with a
given distortion ${\bf R}$ at a point ${\bf r}$ at time $t$. The last
term in equation (\ref{gevola}) contains the thermal fluctuations and
elastic contributions. Here, we have introduced the average velocity
${\bf v}_R=g^{-1} \int {\bf u} f d{\bf u}$. The term $\mathbb{C}_2
\cdot ({\bf v}_0 -{\bf v}_R)$ represents a drag force on the
molecule. We will assume that it is proportional to ${\bf R}$ times the
characteristic stress given by the velocity gradient imposed by
the heat bath. Thus, at first order in ${\bf R}$,
one has $\mathbb{C}_2 \cdot ({\bf v}_R -{\bf v}_0) \simeq -
\mathbb{C}_2 \cdot {\bf R} \cdot \nabla{\bf v}_0$. Notice that
averaging equation (\ref{gevola}) over ${\bf r}$,  an
equation for the distortion vector distribution resembling known
equations for dilute suspensions of polymers is obtained
\cite{bird-hassager}.

From equation Eqn. (\ref{gevola}), we 
can compute the evolution equation for the average distortion field 
$\langle {\bf R} \rangle ({\bf r},t)$
\begin{equation}
\label{Rprom}
\langle {\bf R} \rangle ({\bf r},t) = \frac{1}{n} \int {\bf R} g d{\bf 
R},
\end{equation} 
and for the second moment field $\mathbb{A}({\bf r},t) $ 
\begin{equation}
\label{A}
\mathbb{A} ({\bf r},t)= \langle {\bf RR} \rangle = \frac{1}{n} \int{\bf 
R} {\bf R}
g d{\bf R},
\end{equation} 
which can be interpreted as the moment of inertia tensor of the
molecule \cite{rallison1}. In these equations, $n({\bf r},t) = \int
g\,d{\bf R}$ is the particle density satisfying the continuity
equation $\frac{\partial n}{\partial t} = -\nabla \cdot \left[ n {\bf
v}({\bf r},t) \right]$, obtained by averaging
(\ref{gevola}) over ${\bf R}$. Taking the time
derivative of Eqn.  (\ref{Rprom}) and using (\ref{gevola}), after
integrating by parts one obtains
\begin{eqnarray}\label{Revol}
\frac{d \langle {\bf R} \rangle} {d t} = \frac{1}{n} \nabla \cdot 
\left[n
(\mathbb{A}-\langle {\bf R} \rangle \langle {\bf R} \rangle) \cdot 
\nabla {\bf
v}_0 \right]+ \mathbb{C}_2 \cdot \langle {\bf R} \rangle 
\cdot \nabla {\bf 
v}_0 - \mathbb{L}_{RR} \cdot \langle {\bf F} \rangle,
\end{eqnarray}
where $ \langle {\bf F} \rangle = n^{-1}\int {\bf F} g d{\bf R}$ is
the average elastic force depending on ${\bf r}$ and $t$ through the
average value $\langle {\bf R} \rangle ({\bf r},t)$. The first term in
the right-hand side of Eqn. (\ref{Revol}) contains the standard
deviation $(\mathbb{A}-\langle {\bf R} \rangle\langle {\bf R}
\rangle)$ of the ${\bf R}$ distribution. In the case of a narrow
distribution this term can be neglected, reducing our equation to the
one obtained in Ref. \cite{hinch}. 

Following a similar procedure, from Eqs. (\ref{gevola}) and (\ref{A}) 
it is possible to obtain the evolution equation for $\mathbb{A}$
\begin{eqnarray}\label{Aevol}
\frac{d \mathbb{A}} {d t} - 2\left[\mathbb{A} \cdot \nabla {\bf v}_0
\right]^s = 2 \left\{\mathbb{L}_{RR} \cdot \left[\frac{k_B T} {m}
\mathbb{I} - \langle {\bf FR} \rangle \right]\right\}^s \nonumber \\
+2 \left[ \mathbb{L}_{Rr} \cdot \left( \nabla {\bf v}_0 \cdot \nabla
{\bf v}_0^T \right) \cdot \mathbb{A} \right]^s
\end{eqnarray}  
where the super index $s$ stands for the symmetric part of a tensor,
$\mathbb{I}$ is the identity tensor and we have chosen
$\mathbb{L}_{Ru}=\mathbb{I}$ in (\ref{c1}), to obey the material frame
indifference condition for (\ref{Aevol}), \cite{bird-hassager}. Thus,
the left-hand side of this equation is an Oldroyd time derivative. The
first term in the right-hand side represents the elastic contribution,
comprising the thermal fluctuations of the monomers
($\frac{k_BT}{m}$), and their spatial restrictions due to the fact
that they constitute a chain ($\langle {\bf FR} \rangle$). The last
term in Eqn. (\ref{Aevol}) is a genuine nonlinear contribution in
$\nabla {\bf v}_0$ arising from contracting the description from the
mesoscopic $({\bf r},{\bf u},{\bf R})$-space to the physical ${\bf
r}$-space. Moreover, in this equation, we have neglected contributions
arising from higher order moments of the distribution since their
characteristic relaxation times are at least one order of magnitude
smaller than the one corresponding to $\langle {\bf R} \rangle$ and
$\mathbb{A}$ (see, for example, \cite{checoloco}).

\subsection{Evolution equation for the pressure tensor.}

At hydrodynamic level, the presence of polymer molecules modifies the 
stress field of the flow by adding stresses due to elastic and Brownian 
motion fluctuations. 
These contributions can be calculated by averaging our description over 
the instantaneous velocity ${\bf u}$ and the distortion vector ${\bf 
R}$ of the polymer molecules and using the definitions 
\begin{equation}
\label{presionk}
\mathbb{P}^K=m\int({\bf u}-{\bf v})({\bf u}-{\bf v})fd{\bf u}d{\bf R},
\end{equation}
and
\begin{equation}
\mathbb{P}^E=-\frac{m}{2}\int({\bf
FR}+{\bf RF})fd{\bf u}d{\bf R},
\end{equation}
where $\mathbb{P}^K$ and $\mathbb{P}^E$ are the kinetic and elastic 
contributions of the suspended phase to the total pressure tensor, 
respectively, \cite{mclennan,kirkwood}. 

The equations for the pressure tensor contributions can be
obtained from the evolution equations for the moments of the
probability density $f$ and the Fokker-Planck equation
(\ref{fuckingplanck}). Contributions from third and higher order
moments of the hydrodynamic hierarchy have been neglected
\cite{checoloco}. For the kinetic contribution, using Eqns. 
(\ref{fuckingplanck}), (\ref{c1}) and (\ref{presionk}), one obtains the evolution 
equation 
\begin{equation}
\label{evolpresionk}
\frac{d \mathbb{P}^K}{d t} +2 \left[ \mathbb{P}^K \cdot \left(
\mathbb{C}_1 + \nabla {\bf v}_0 \right) \right]^s = 2 \frac{k_B T}{m}
\rho (\mathbb{C}_1-\mathbb{L}_{ur} \cdot  \nabla {\bf v}_0)^s.
\end{equation}
The time derivative in equation (\ref{evolpresionk}) accounts for
memory effects in a similar fashion as Maxwell like models do
\cite{bird-hassager}. For times $t \gg \left[ \mathbb{C}_1 + \nabla {\bf v}_0
\right]^{-1}_{ij}$, the time derivative in equation
(\ref{evolpresionk}) can be neglected, giving an algebraic system of
equations for the components of $\mathbb{P}^K$ that can be solved for
any imposed flow. Eqn. (\ref{evolpresionk}) incorporates contributions
to the stresses arising from the flow inhomogeneities through $ \nabla
{\bf v}_0$. These contributions have not been considered in previous
descriptions where Brownian stresses were assumed to be given by
$\mathbb{P}^K \simeq \frac{k_B T}{m} \rho \mathbb{I}$,
\cite{bird71}. To simplify the computation of the elastic contribution
to the pressure tensor, in the following we will consider that the
elastic force ${\bf F}$ is proportional to ${\bf R}$ with a
proportionality coefficient $\xi_0 \tilde{F}$, where $\tilde{F}$ is a
dimensionless function of $tr(\mathbb{A})$ giving the form
of the spring law, and  $\xi_0$ is the characteristic spring restitution
coefficient per unit mass. Therefore, the elastic
contribution is $\mathbb{P}^E=-\xi_0 \tilde{F} \rho \mathbb{A}$ and
its evolution is related to equation (\ref{Aevol}).

We will now proceed to express the evolution equations (\ref{Aevol})
and (\ref{evolpresionk}) in dimensionless form. In the absence of an
externally imposed flow, equation (\ref{Aevol}) suggests that
$\mathbb{A}$ scales with $k_{B} T/ m \xi_0$.  Therefore, we will scale
length with $\sqrt{k_B T/4 m \xi_0}$, time with the inverse of the
shear rate $\dot{\gamma}$ and pressure with $\eta_s \dot{\gamma}$,
where $\eta_s$ is the solvent viscosity.  Since the Onsager
coefficient $\mathbb{L}_{RR}$ represents the mobility, it will be
scaled with the inverse of the characteristic friction coefficient per
unit mass, say $\beta$, \cite{warner}. The coefficient
$\mathbb{L}_{Rr}$, will be scaled with $\beta/\xi_0$ as it charactizes
the coupling between the distortion of the molecule and the drag
forces, see Eq. (\ref{jr}). Thus, the evolution equation (\ref{Aevol})
takes the dimensionless form
\begin{eqnarray}\label{Aevol-Adim}
\frac{d \mathbb{A}} {d t} - 2\left[\mathbb{A} \cdot \nabla {\bf v}_0 
\right]^s = \frac{2}{D}\left[ \mathbb{L}_{RR} \cdot
\left( \mathbb{I} - \tilde{F} {\mathbb{A}} \right)\right]^s \nonumber
\\ + 2 D 
\left[ \mathbb{L}_{Rr} \cdot  \left( \nabla {\bf v}_0 \cdot \nabla {\bf 
v}_0^T \right) \cdot \mathbb{A}  \right]^s,
\end{eqnarray}  
where $D=\dot{\gamma}\beta/\xi_0$ is the Deborah number. In similar
form, the dimensionless evolution equation for $\mathbb{P}^K$ is given
by
\begin{equation}
\label{evolpresionk2}
\frac{d \mathbb{P}^K}{d t} + 2\left(
\mathbb{P}^K \cdot\nabla {\bf v}_0 \right)^s
 = \frac{2c}{D}
\left( \frac{D^*}{D} \mathbb{C}_1-\mathbb{L}_{ur} \cdot
\nabla {\bf v}_0 \right)^s-2 \frac{D^*}{D} \left(\mathbb{P}^K \cdot 
\mathbb{C}_1 \right)^s,
\end{equation}
where $c=\rho k_B T \beta/ \eta m \xi_0$ is a measure of the
concentration of polymers, $D^*=\beta^2/\xi_0$ is a dimensionless
parameter that
compares viscous dissipation and molecule relaxation times. We have
scaled the friction tensor $\mathbb{C}_1$ with $\beta$. Finally,
notice that the dimensionless elastic contribution to the pressure
tensor is
\begin{equation}
\label{presione2}
\mathbb{P}^E=-\frac{c\tilde{F}}{2D} \mathbb{A}.
\end{equation}

Eqns. (\ref{Aevol-Adim}), (\ref{evolpresionk2}) and (\ref{presione2}) 
represent the constitutive relations for a "gas" of deformable particles 
embedded into a Newtonian heat bath. It is worth noticing that in this 
simple model we have characterized the internal dynamics of the 
particle by using the single vector parameter ${\bf R}$, known as the gross 
distortion of the particle. 
In the following section, we will apply these equations to the case of 
a simple shear flows with different models for the restoring force of 
the particles and for the Onsager coefficients. 

\section{Applications}

We will now proceed to consider the constitutive relation expressed in
terms of the evolution equations (\ref{Aevol-Adim}) and
(\ref{evolpresionk2}) for particular forms of the Onsager
coefficients, corresponding to simplified models of the suspended
polymer molecules under shear flow conditions. With these
simplifications, the constitutive relation reduces to models that, at
first order in $\nabla {\bf v}_0$, resemble the known FENE models for
polymer solutions. 

The cases studied were a dumbbell, a dumbbell with an isotropic
friction coefficient dependent on its gross distortion, and a dumbbell
with a non-isotropic friction also dependent on the gross distortion
of the molecule. Solutions are compared with experiments.

{\it Dumbbells in a simple shear}.  In this case, the dimensionless
Onsager coefficients $\mathbb{C}_1$, $\mathbb{L}_{RR}$ and
$\mathbb{L}_{ur}$ can be assumed to be isotropic and constant. In
particular we will assume that $\mathbb{C}_1 = \mathbb{L}_{RR} =
\mathbb{I}$ and neglect non-linear terms in $\nabla {\bf v}_0$ by
taking $\mathbb{L}_{Rr} = 0$. For spheres,
\begin{equation}
\label{lur}
\mathbb{L}_{ur}= \frac{ma^2 \beta^2}{6 k_B T}  \mathbb{I} ,
\end{equation}
where $a$ is the radius of the sphere, \cite{checoloco2}. Taking the
radius 
$a$ as the characteristic length scale $\sqrt{k_B T/4 m \xi_0}$ of the 
particle one obtains $\mathbb{L}_{ur}= \frac{D^*}{24} \mathbb{I} $. Additionally, we 
will adopt the FENE spring law \cite{warner}
\begin{equation}
\label{f}
\tilde{F}=\frac{L^2}{L^2- tr(\mathbb{A})}.
\end{equation}
For times $t \gg \left[ \mathbb{C}_1 + \nabla {\bf v}_0
\right]^{-1}_{ij}$, the time derivative in equations (\ref{Aevol-Adim})
and (\ref{evolpresionk2}) can be neglected, leading to a set of
algebraic equations for $\mathbb{P}^k$ and $\mathbb{A}$. Therefore, 
$\mathbb{A}$ is a solution of a non-linear algebraic equation and its 
components are given by:
$A_{xx} = L^2(1-A_{yy})-2A_{yy}$, $A_{xy} = 2 A_{yy}^2 D$ and
$A_{zz} = A_{yy}$. With $A_{yy}$ the real solution of 
\begin{equation}
\label{cubica}
8D^2 A_{yy}^3 + (3+L^2)A_{yy} -L^2=0.
\end{equation}
The behavior of $A_{xx}$ and $A_{yy}$ as functions of $D$ are common
to FENE models \cite{bird-hassager}. For large values of $D$, $A_{xx}$ and
$A_{yy}$ will tend to $L^2$ and zero respectively. The solution to the
elastic pressure tensor $\mathbb{P}^E$ depends on $\mathbb{A}$ and has
no compact expression, therefore it is not shown here.
Similarly, the components of the kinetic contribution to the pressure
tensor are given by
\begin{equation}
\label{pkxy} 
P_{xy}^K=- \frac{c}{2} \left( \frac{1}{24}+\frac{1}{D^*}  \right)
\end{equation}
and
\begin{equation}
P_{xx}^K-P_{yy}^K=\frac{cD}{2D^*} \left( 
\frac{1}{24}+\frac{1}{D^*}\right),
\end{equation}
whereas the dimensionless first normal stress difference is defined by
\begin{equation}
\label{pkxx-pkyy} 
N_1=-D \left( P_{xx}^K-P_{yy}^K+P_{xx}^E-P_{yy}^E \right).
\end{equation}
The behavior of $N_1$ as a function of $D$ reproduce the quadratic
response of polymer solutions at low Deborah numbers observed in FENE
models \cite{chilcott,chhabra}. We have obtained two contributions to the shear
stress: $\sigma= -D(P_{xy}^K+P_{xy}^E)$. The contribution of the
elastic part is similar to that of the usual FENE-P models
\cite{bird-hassager}. A new contribution arises from the kinetic part
and corresponds to that of a 'gas' of Brownian particles
\cite{checoloco}.

{\it Isotropic non-constant friction}.
In this case, the interaction between the molecule and the bath is
modeled through a size dependent friction coefficient. We are assuming
that the average shape of the molecule ranges from a sphere in
quiescent liquid to a symmetrical spheroid with a principal axis $L$
in the case of maximum distortion.  In such case, the friction
coefficient $\mathbb{C}_1$ will be a tensorial quantity. However, the
ellipsoid will be considered to be aligned with the flow and, under
this condition, the friction coefficient will be taken to be that of a
spheroid in an axisymmetric unbounded flow given by 
\begin{equation}
\label{beta}
\beta=\beta_0 
\frac{\sqrt{a^2-b^2}}{(\alpha^2+1)\coth^{-1}\alpha-\alpha},
\end{equation}
where $\beta_0$ is now the Stokes friction coefficient, $a$ the
principal spheroid axes along the flow, $b$ the perpendicular one and
$\alpha =a/ \sqrt{a^2-b^2}$, \cite{happel}. Since in our formulation the molecule
deformation is solely given by ${\bf R}$, then $a
\simeq \frac{1}{2}|{\bf R}|=\frac{1}{2}\sqrt{tr(\mathbb{A})}$. Hence,
the maximum value of $a$ is $\frac{1}{2} L$ and as a consequence,
$\alpha$ ranges form $L/\sqrt{L^2-4b^2}>1$ in the case of maximum
deformation, to infinity in the case of a sphere (no
deformation). From this consideration, it follows that
$\coth^{-1}\alpha$ is never singular. Then, approximating $\coth^{-1} \alpha \simeq
\frac{1}{\alpha}+\frac{1}{3 \alpha^3}$, equation 
(\ref{beta}) becomes
\begin{equation}
\label{beta2}
\beta \simeq \frac{9}{32}  \beta_0 \sqrt{ tr(\mathbb{A})} \left[
  1+\frac{4b^2}{3 tr(\mathbb{A})} \right].
\end{equation} 
Notice that,
by definition, $b$ must be a decreasing function of $tr(\mathbb{A})$,
therefore, the second term in the approximation (\ref{beta2}) must be
negligible small as soon as the molecule model begins to experience 
some distortion. 
Now, by using Eqn. (\ref{beta2}) the dimensionless evolution 
equation for the distortion tensor $\mathbb{A}$ takes the form
\begin{equation}\label{Aevol2-spheroid}
\frac{d \mathbb{A}} {d t} - \mathbb{A} \cdot  \nabla {\bf v}_0
-  \nabla {\bf v}_0^T 
\cdot \mathbb{A} = \frac{16}{9 \sqrt{ tr(\mathbb{A})}}\frac{1}{D}
\left(\mathbb{I} - \tilde{F} \mathbb{A} \right),
\end{equation}
which constitutes a correction to the friction equivalent to those
used in Refs. \cite{degennes,rallison1,hinch}. Notice that the
dependence of the friction coefficient on $\mathbb{A}$ introduces a
non-linear coupling between Eqn. (\ref{Aevol2-spheroid}) and the
corresponding evolution equation for $\mathbb{P}^K$.
Solutions for $\mathbb{A}$ and $\mathbb{P}^K$ can be obtained by
following the procedure of the previous subsection. 

\begin{figure}[tt]
\begin{center}	
\includegraphics[width=0.8\textwidth]{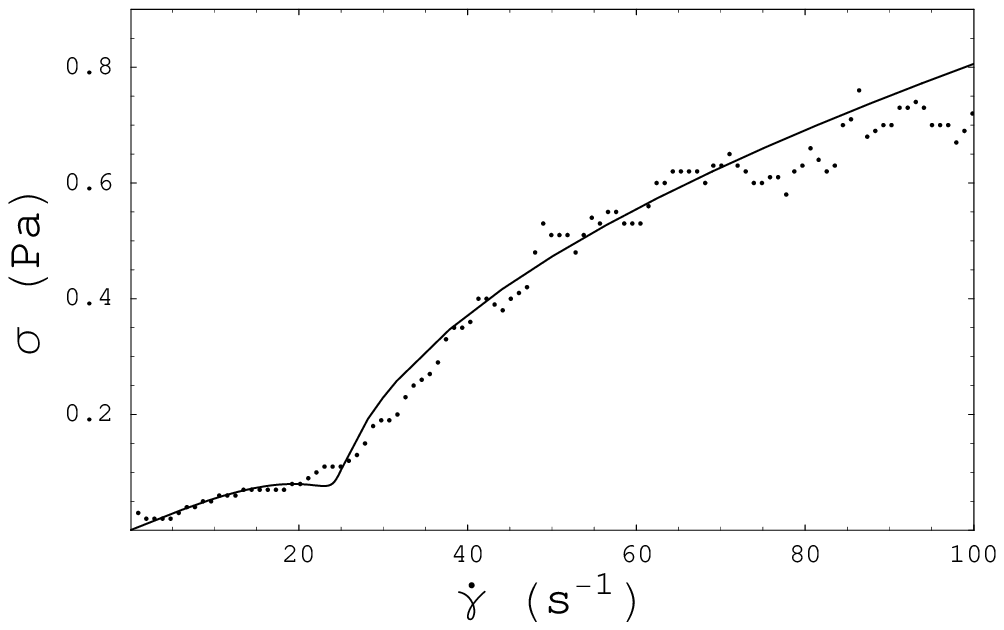}
\caption{Fit of experimental data (points) for the non-Newtonian shear stress
  versus the shear rate, using the isotropic friction model (line). The fit is
  made using
 $c \eta = 0.0038 kg/m s$, $\beta=3.5 s^{-1}$,
  $\xi=17 s^{-2}$ and $L=80$.}
\label{iso}
\end{center}
\end{figure}

Figure (\ref{iso}) shows a fit of the {\it isotropic} friction model to
simple shear experiments with dilute micellar solutions in water
\cite{jorge}. Experiments were performed increasing the shear rate
$\dot{\gamma}$ from zero in steps, the steps in shear rate were taken
at an acceleration of $0.66 s^{-2}$. For shear rates over $100 s^{-1}$
the model overpredicts stresses. The kinetic contribution is
responsible for the agreement for $\dot{\gamma}$ below $25 s^{-1}$
whereas the elastic contribution dominates for larger shear rates.

{\it Non-isotropic friction}. Here we will take the friction tensor
$\mathbb{C}_1$ to be diagonal but non-isotropic. This case represents
a more realistic model than the previous ones because it better
accounts for the drag on a body with cylindrical symmetry oriented with
the flow, \cite{happel}. Thus, in a shear flow on the $x$ direction
$\mathbb{L}_{RR}=\mathbb{C}_{1}^{-1}$,
$\mathbb{L}_{ur}= \frac{ma^2}{6 k_B T} \mathbb{C}_1 \cdot
\mathbb{C}_1$ and $\mathbb{L}_{Rr}=0$ for simplicity.
Where the friction tensor will be modelled by 
\begin{equation}
\mathbb{C}_{1}=\beta_0
 \left[
  \begin{array}{ c c c}
     \frac{9}{32} \sqrt{tr(\mathbb{A})} & \,\,\, 0 & \,\,\,\,\,\,\, \,\,\, \,\,\,0 \\
     0 &\,\,\, 1 & \,\,\,\,\, \,\,\, \,\,\,\,\,0 \\
     0 & \,\,\,0 & \,\,\,\,\, \,\,\, \,\,\,\,\,1 
  \end{array} \right].
\end{equation}

\begin{figure}[tt]
\begin{center}	
\includegraphics[width=0.8\textwidth]{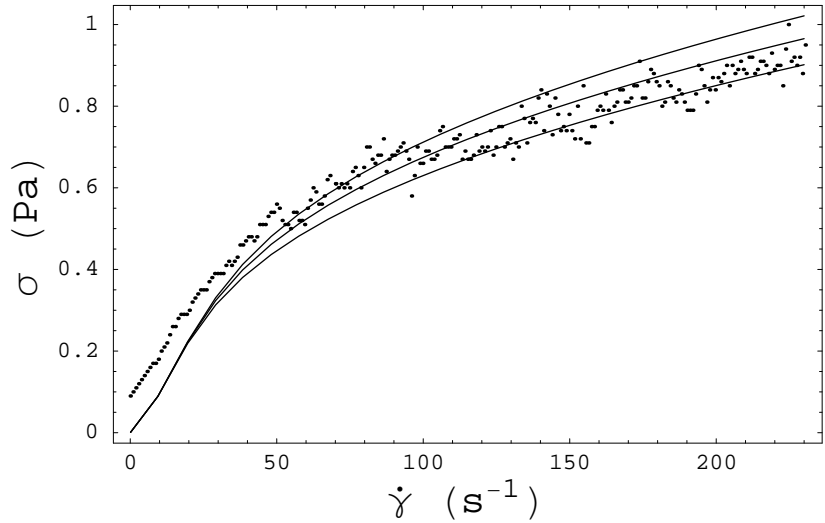}
\caption{Fit of experimental data (points) for the non-Newtonian shear
  stress versus the shear rate, using the non-isotropic friction model
  (lines). The fit is made using $c \eta = 0.025 kg/m s$, $\beta=5
  s^{-1}$, $\xi=11 s^{-2}$ and $L=110,140,170$ starting from bottom.}
\label{aniso}
\end{center}
\end{figure}

Figure (\ref{aniso}) shows a fit of the {\it non-isotropic} friction
model to experiments with the same system \cite{jorge} but in this
case decreasing the shear rate $\dot{\gamma}$ to zero in steps at the
same acceleration. For this model elastic stresses dominate at all
shear rates. The models suggests that the difference between
experiments arises from the possible configurations that the particles
can explore in each case. The {\it isotropic} friction model best fits
the stress behavior in the coil-stretch transition that occurs during
shear rate growth. The residual stress at zero shear rate obtained in
experiments (Figure (\ref{aniso})) suggests that particles have not
reached a coiled configuration. This explains why the
{\it non-isotropic} friction model reproduces the stress behavior in
shear rate reduction.

Notice that, in this description, the dynamics of the molecule
segments is not included. Therefore, the effect of hydrodynamic
interactions between segments are not contained explicitly in the set
of Onsager coefficients appearing in equations (\ref{ju}) and
(\ref{jr}). If considered, these interactions will contribute to the
rheology of the model by modifying the constitutive relations through
a different set of Onsager coefficients. The aim of the present model
was to estimate the collective effect of the molecule segments through
a set of effective Onsager coefficients \cite{doi}.

\section{Conclusions}

A novel approach to the description of the dynamics of dilute polymer 
solutions has been explored using the formalism of mesoscopic 
nonequilibrium thermodynamics in the simplified case of a single 
vector parameter characterizing the polymer configuration, common to 
FENE models.

The description has been carried out by formulating the laws of
nonequilibrium thermodynamics after assuming valid the hypothesis of
local equilibrium in phase space. By calculating the entropy
production of the system, we have obtained general constitutive
equations at mesoscopic level for the dissipative currents in ${\bf
r}$, ${\bf R}$ and ${\bf u}$ spaces. Using then probability
conservation, a generalized Fokker-Planck equation describing the
dynamics of the system has been obtained.

A general closure to the governing equations for the hydrodynamics of
this system model is derived from the Fokker-Planck equation. Those
equations consist in two evolution equations for elastic and kinetic
contributions to the pressure tensor. The evolution equation for the
elastic stresses resembles known FENE models. The kinetic
contribution, having no counterpart in the FENE models, enters into
the hydrodynamic description due to the coupling between elastic and
Brownian effects in the mesoscopic constitutive relations.

This method can generate a family of constitutive relations in
configuration space depending on the modeled Onsager
coefficients. Three examples were analyzed in the case of simple shear
flow. In the first one, we have analyzed the dumbbell model with
constant friction recovering the elastic behavior found in the FENE
models \cite{chilcott}. In the second case, the friction of the
molecule has been modeled with that of a spheroid aligned with the
flow \cite{degennes,rallison2}. In this case we have obtained a
steeper coil-stretch transition and good agreement with experiments of
shear rate growth. In the last example we have considered the
tensorial character of the Onsager coefficients by taking the
non-isotropic friction tensor of a spheroidal particle aligned with
the flow. This model is capable of describing the shear stress
behavior in shear rate reduction experiments.

Given that the general constitutive equations here obtained can be
reduced, with simple arguments, to models that predict experimental
data, we are confident that further analysis can provide simple
mathematical expressions for constitutive relations better approaching
more complex situations. For example, in the case where
configurational rates of change are relevant, the time change in the
gross distortion of the molecule ought to be included in the
description of the system, introducing an additional relaxation time.

\section{Acknowledgments}

We wish to acknowledge J. M. Rub\'{\i}, A. P\'{e}rez-Madrid,
I. Pagonabarraga for their valuables comments and
discussions, we also want to thank R. Peralta-Fabi for
his comments. This work was partially supported by UNAM-DGAPA.



\newpage

\newpage

\end{document}